\documentclass[lettersize,journal]{IEEEtran}
\usepackage{color,xcolor}
\usepackage{epsfig}
\usepackage{graphicx}
\usepackage{lipsum}
\usepackage{pifont}
\usepackage{array}
\usepackage{booktabs}
\usepackage{colortbl}
\usepackage{multirow}
\usepackage{float}
\usepackage{caption}
\usepackage[labelformat=simple]{subcaption}
\usepackage{adjustbox}

\usepackage{footnote}
\makesavenoteenv{tabular}
\makesavenoteenv{table}

\usepackage{amsmath,amsfonts,amssymb,bm}
\usepackage[super]{nth}

\usepackage{changepage}
\usepackage{extramarks}
\usepackage{fancyhdr}
\usepackage{lastpage}
\usepackage{setspace}
\usepackage{soul}
\usepackage{xspace}

\usepackage{url}
\usepackage{hyperref}
\hypersetup{colorlinks=True, urlcolor=black}
\usepackage[numbers,sort&compress]{natbib}
\usepackage{algorithm, algorithmic}
\usepackage{enumitem}
\usepackage{verbatim}
\usepackage{pifont}
\usepackage[acronyms]{glossaries}
\glsdisablehyper

\newcommand{\sectdot}[1]{Sec.~\ref{sec:#1}}

\newcommand{\eqndot}[1]{Eqn.~(\ref{eqn:#1})}

\newcommand{\figdot}[1]{Fig.~\ref{fig:#1}}
\newcommand{\tbl}[1]{Table~\ref{tab:#1}}

\newcommand{\twosectdot}[2]{Sec.~\ref{sec:#1} and \ref{sec:#2}}

\newcommand{\twotbl}[2]{Tables~\ref{tab:#1} and \ref{tab:#2}}

\newcommand{\ignore}[1]{}

\makeatletter
\DeclareRobustCommand\onedot{\futurelet\@let@token\@onedot}
\def\@onedot{\ifx\@let@token.\else.\null\fi\xspace}

\makeatother

\definecolor{MyDarkBlue}{rgb}{0,0.08,1}
\definecolor{MyDarkGreen}{rgb}{0.02,0.6,0.02}
\definecolor{MyDarkRed}{rgb}{0.8,0.02,0.02}
\definecolor{MyDarkOrange}{rgb}{0.40,0.2,0.02}
\definecolor{MyPurple}{RGB}{111,0,255}
\definecolor{MyRed}{rgb}{1.0,0.0,0.0}
\definecolor{MyGold}{rgb}{0.75,0.6,0.12}
\definecolor{MyDarkgray}{rgb}{0.66, 0.66, 0.66}

\def\presec{\vspace{-0.2em}}
\def\postsec{\vspace{-0.2em}}
\def\pretbl{\vspace{-0.2em}}
\def\posttbl{\vspace{-0.5em}}

\usepackage{pgfplots}
\DeclareUnicodeCharacter{2212}{−}
\usepgfplotslibrary{groupplots,dateplot}
\usetikzlibrary{patterns,shapes.arrows}
\pgfplotsset{compat=newest}
\begin{document}

\title{Knowledge Distillation from Multiple Foundation Models for End-to-End Speech Recognition}

\author{Xiaoyu Yang,~\IEEEmembership{Student Member,~IEEE,}, Qiujia Li,~\IEEEmembership{Member,~IEEE,}\\Chao Zhang,~\IEEEmembership{Member,~IEEE,} Philip C. Woodland,~\IEEEmembership{Fellow,~IEEE}

\thanks{This paper was produced by the IEEE Publication Technology Group. They are in Piscataway, NJ.}}

\markboth{SUBMITTED TO IEEE/ACM TRANSACTIONS ON AUDIO SPEECH AND LANGUAGE PROCESSING}%
{Shell \MakeLowercase{\textit{et al.}}: A Sample Article Using IEEEtran.cls for IEEE Journals}


\maketitle
\begin{abstract}
Although large foundation models pre-trained by self-supervised learning have achieved state-of-the-art performance in many tasks including automatic speech recognition (ASR), knowledge distillation (KD) is often required in practice to transfer the knowledge learned by large teacher models into much smaller student models with affordable computation and memory costs. This paper proposes a novel two-stage KD framework to distil the knowledge from multiple speech foundation models as teachers into a single student neural transducer model for ASR. In the first stage, the student model encoder is pre-trained using the embeddings extracted from multiple teacher models. In the second stage, the student encoder is fine-tuned with the audio-text pairs based on the ASR task. Experiments on the LibriSpeech 100-hour subset show that the proposed KD framework improves the performance of both streaming and non-streaming student models when using only one teacher. The performance of the student model can be further enhanced when multiple teachers are used jointly, achieving word error rate reductions (WERRs) of 17.5\% and 10.6\%. Our proposed framework can be combined with other existing KD methods to achieve further improvements. Further WERRs were obtained by incorporating extra unlabelled data during encoder pre-training, leading to a total relative WERR of 55.0\% on the non-streaming student model.
\end{abstract}

\begin{IEEEkeywords}
Foundation model, teacher-student training, multi-teacher knowledge distillation, neural transducer, ASR
\end{IEEEkeywords}
\section{Introduction}

\IEEEPARstart{A}{utomatic} speech recognition (ASR) is the task of mapping an input speech signal to its corresponding text sequence. 
An end-to-end (E2E) trainable ASR model implements this using a single trainable neural network such as neural transducers~\cite{graves2012RNNT,Graves2013} and attention-based encoder-decoder models~\cite{Bahdanau2015EndtoendAL,Lu2015,chan2015listen,Kim2017}.  
As a result, E2E trainable models have become the most prevalent ASR approach in both industry and academia. 

In real-world applications, the size of an E2E trainable ASR system is usually limited by computation and memory budgets. Knowledge distillation (KD)~\cite{hinton2015knowledgedistillation}, also known as teacher-student training, is a commonly used approach for model compression, 
from the original large model ("teacher") to  the compressed small model ("student").
Instead of training the student purely using ground truth labels, KD also enables the use of the outputs of the teacher model, namely teacher labels. 
Different KD loss functions and training targets for KD to improve the performance of the student ASR model have been previously explored. For example, Takashima et al~\cite{takashima2018investigation} uses the final output distributions as the teacher labels and the Kullback-Leibler (KL) divergence as the KD loss together with the connectionist temporal classification (CTC) loss for ASR. 
In \cite{swaminathan2021codert} hidden layer representations or embeddings were extracted as the teacher labels for training the student ASR model. Unlike the parameter quantisation-based methods~\cite{hwang2014fixed, courbariaux2015binaryconnect} that usually require the teacher and student to share the same model structure, KD does not impose this constraint, making it a more flexible approach. Meanwhile, as ground-truth labels are not required, KD can be used as a semi-supervised training method to leverage a large amount of unlabelled data to improve the student~\cite{menghani2019learning,Li2017LargeScaleDA}. 
Furthermore, the student model can also be jointly trained using multiple teachers. Since the ensemble of multiple models can usually outperform a single model~\cite{zhou2021ensemble}, it can serve as a more reliable teacher during KD training. 
The weighted sum of the distributions output from multiple teachers is used as distillation targets in \cite{ensembledistribution}, leading to a larger performance gain compared to using only a single teacher. Wong et al~\cite{JeremyDecisionTree} ensembled the teachers with different output nodes obtained by decision tree clustering, and sequence-level KL divergence was also developed as an improved KD loss~\cite{Wong2016SequenceST}.  
In \cite{fukuda2017efficient}, one teacher from a teacher pool was selected in every mini-batch to generate the teacher labels.

Recently, pre-trained foundation models~\cite{bommasani2021foundation} obtained by self-supervised learning (SSL)~\cite{SSLloss} have achieved state-of-the-art performance on many speech processing tasks~\cite{baevski2020wav2vec,hsu2021hubert, chen2022wavlm,chen2020big,wang2021unispeech,kreyssig2022biased}. After being pre-trained on a large amount of unlabelled speech data, foundation models can be fine-tuned with the labelled data for specific downstream tasks, such as ASR~\cite{baevski2020wav2vec, hsu2021hubert}, speaker diarisation~\cite{chen2022wavlm,zheng2022tandem}, and emotion recognition~\cite{pepino2021emotion,Shen2020}  \textit{etc}. However, foundation models tend to be very large (e.g. from hundreds of millions~\cite{baevski2020wav2vec, hsu2021hubert} to even billions~\cite{hsu2021hubert} of model parameters) in order to fully leverage the richness and diversity of the unlabelled data during pre-training. This prevents their uses in resource-constrained scenarios, such as on-device streaming ASR, due to the large latency, computational footprint and inference cost. Therefore, it is of great interest to compress these large foundation models with as little performance degradation as possible. To this end, Peng et al~\cite{peng2021shrinking} projected the embeddings generated by a wav2vec 2.0 model~\cite{baevski2020wav2vec} to distributions and performed KD based on the KL divergence. Chang et al~\cite{chang2022distilhubert} proposed a multi-task KD framework by predicting embeddings extracted from different hidden layers of a HuBERT~\cite{hsu2021hubert} layer with multiple heads.

Due to use of various loss functions, model structures, and training data, embeddings generated by different foundation models can be very different and complementary. Therefore, further potential performance improvement could be achieved if a student model can learn from multiple foundation models. 
In this paper, we propose an efficient 2-stage distillation method to train student neural transducers for E2E ASR using the knowledge distilled from multiple foundation models as teachers. KD is carried out first at the embedding level and then at the hypothesis level. At the embedding level, the student jointly learns the embeddings from multiple teacher models using a regression loss. At the hypothesis level, the top two ASR hypotheses obtained by beam search (termed as \textit{1-best} and \textit{2-best}) can be used as the supervision in the transducer loss. 
The main contributions of this paper are summarised as follows:
\begin{enumerate}
    \item Proposes a two-stage KD framework which is effective both for streaming and non-streaming transducer models;
    \item Demonstrates that using multiple teachers in the proposed KD framework leads to lower student word error rates (WERs) compared to a single teacher;
    \item Shows that the proposed KD framework is complementary with existing KD methods for neural transducers, and lower student WERs can be achieved when incorporating more unlabelled data.
\end{enumerate}

In the rest of this paper, \sectdot{related} briefly reviews the background and related work.  
In \sectdot{multi-teacher KD}, details of the proposed two-stage KD framework are presented. 
The experimental setups and results are given in \twosectdot{exp_setup}{exp_results}. Finally, conclusions are drawn in \sectdot{conclusions}.

\section{Related work}
\label{sec:related}

\subsection{Neural transducer models}
\begin{figure}
    \centering
    \includegraphics[width=0.75\linewidth]{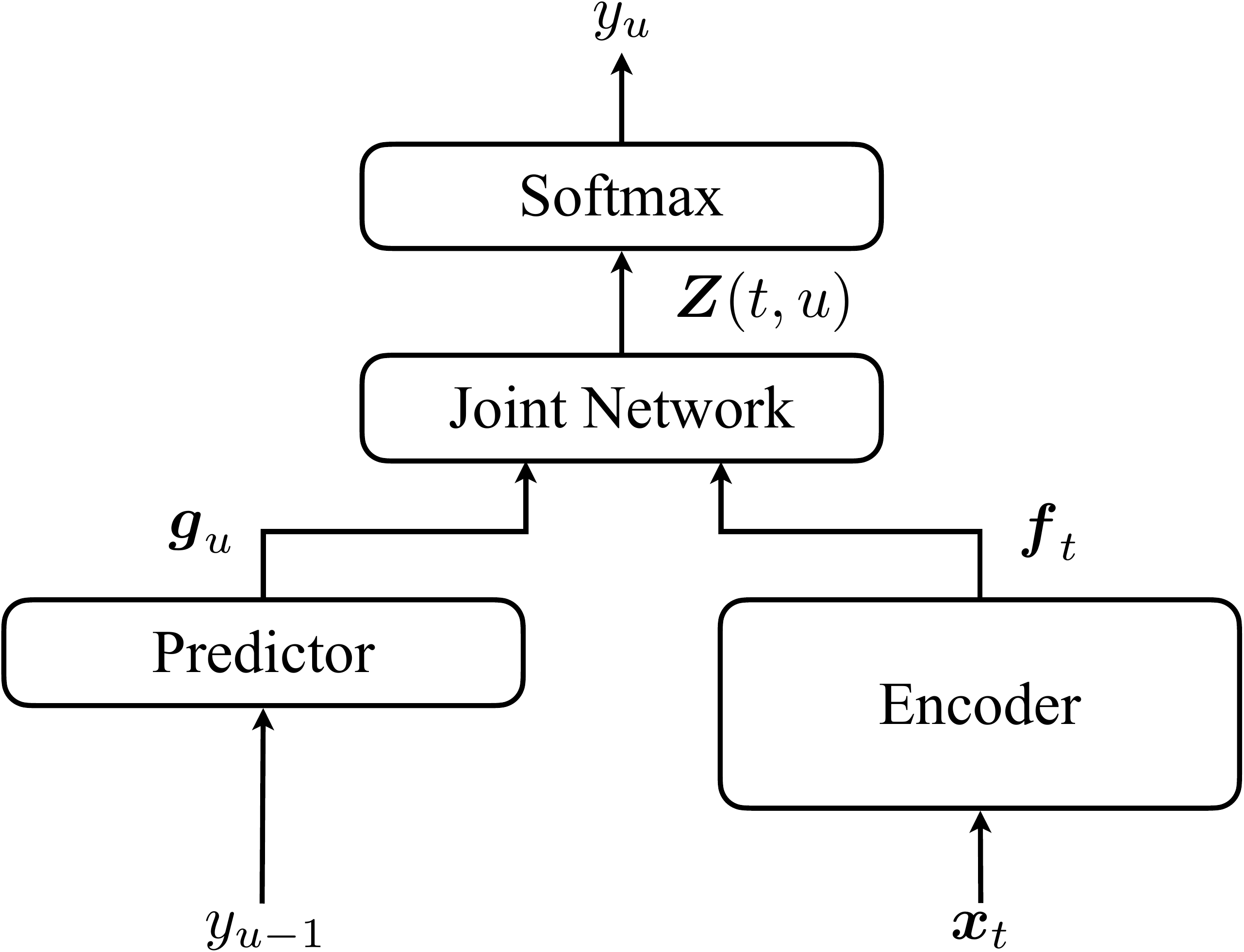}
    \caption{A neural transducer model.}
    \label{fig:rnnt}
\end{figure}

The neural transducer \cite{graves2012RNNT, Graves2013} is a prevalent E2E trainable ASR approach. It consists of an encoder, a predictor and a joint network (see \figdot{rnnt}). Given a pair of input feature sequence $\bm{X}=\bm{x}_1,\ldots,\bm{x}_M$ of length $M$ and its transcription $\bm{y}=y_1,\ldots,y_U$ of length $U$, the neural transducer is trained to maximise the conditional probability $P(\bm{y}|\bm{X})$. The encoder receives $\bm{X}$ and generates an acoustic embedding sequence $\bm{F}=\bm{f}_1,\ldots,\bm{f}_T$ of length $T$. In practice, $T$ is often set to $M$, or a value smaller than $M$ by a constant factor (\textit{e.g.} when $\bm{X}$ is a {raw waveform} instead of acoustic feature sequence). The predictor produces a text embedding $\bm{g}_u$ for each text token $y_u$, and $\bm{G}=\bm{g}_1\ldots,\bm{g}_U$. The joint network takes each pair of $\bm{f}_t$ and $\bm{g}_u$ as input and generates a $V$-dimensional (-d) distribution, where $V$ is the size of the output layer (i.e the token vocabulary) including an extra \textit{blank symbol} which is removed from the final output sequence. 

A (transducer) lattice refers to the collection of all valid alignments between $\bm{F}$ and $\bm{y}$ (and thus $\bm{X}$ and $\bm{y}$) traversing from $(t=0,u=0)$ to $(t=T,u=U)$. The training objective $P(\bm{y}|\bm{X})$ is then the sum of the probability of all valid alignments: 
\begin{align}
    P(\bm{y}|\bm{X}) = \sum\nolimits_{\alpha \in \mathcal{A}}P(\alpha|\bm{X})
\end{align}
where $\mathcal{A}$ is the set containing all valid alignments between $\bm{X}$ and $\bm{y}$. Enumerating all valid alignments is computationally prohibitive and the forward-backward procedure can be used~\cite{graves2012RNNT, Graves2013} to efficiently compute the summation (the RNNT loss). During decoding, Viterbi beam search can be applied to generate output tokens in a time-synchronous manner. Compared to other E2E ASR approaches, including connectionist temporal classification (CTC)~\cite{graves2006CTC} attention-based encoder-decoder (AED) models~\cite{Bahdanau2015EndtoendAL,chan2015listen,watanabeHybrid}, the neural transducer generally achieves lower WERs than CTC and more naturally handles streaming input speech than AED.
In addition, we use the term \textit{distribution lattice} to refer to the $(T,U,V)$-d tensor composed by $\bm{F}$, $\bm{G}$ and all output distributions, which is denoted as $\bm{Z}$. Each node in the distribution lattice, $\bm{Z}(t,u,k)$, represents the probability of generating $k$-th token in the vocabulary at time $t$ after generating the partial text sequence $y_{1},\ldots,y_u$. 

\presec
\presec
\subsection{Knowledge distillation for neural transducers}
\postsec

KD~\cite{hinton2015knowledgedistillation} is a widely used technique that compresses a large teacher model to a small student model by training the small model to match the output of the large model. Depending on the output of the teacher model, different loss functions can be used. For example, the KL-divergence is commonly used as the distillation loss to match probability distributions whereas $L_1$ or $L_2$ loss is more appropriate for matching hidden representations. Since the output of a neural transducer for an input acoustic sequence is a distribution lattice, applying KD to neural transducers is inherently expensive. A straightforward way of applying KD to neural transducers is to use the KL-divergence between the teacher distribution lattice $\bm{Z}_{\mathcal{T}}(t,u,k)$ and the student distribution lattice $\bm{Z}_{\mathcal{S}}(t,u,k)$ as the distillation loss:
\begin{align}
    \mathcal{L}_{\text{KD}} = -\sum_{t=1}^{T}\sum_{u=1}^{U}\sum_{k=1}^{V} \bm{Z}_{\mathcal{T}}(t,u,k) \log \bm{Z}_{\mathcal{S}}(t,u,k).
    \label{eqn:fulllatticeKD}
\end{align}

In practice, however, directly applying the KL-divergence loss defined in \eqndot{fulllatticeKD} is inefficient, due to the high computational and storage costs for the distribution lattice. To reduce cost, \cite{panchapagesan2021efficient} kept only three elements in each node of the original distribution lattice: the probability of the correct symbol $y_u$, the probability of the blank symbol and the sum of the probabilities of all other symbols, and used this collapsed distribution lattice for KD:
\begin{align}
    \mathcal{L}_{\text{KD}}\approx -\sum_{t=1}^{T}\sum_{u=1}^{U}\sum_{k=1}^{3} \bm{Z}'_{\mathcal{T}}(t,u,k) \log \bm{Z}'_{\mathcal{S}}(t,u,k)
    \label{eqn:collapsedKD}
\end{align}
where $\bm{Z}'_{\mathcal{T}}(t,u,k)$ and $\bm{Z}'_{\mathcal{S}}(t,u,k)$ are the collapsed versions of the teacher and student posterior lattice.
As a result, the computational complexity is reduced from $\mathcal{O}(TUV)$ to $\mathcal{O}(TU)$. However, a large amount of information is discarded 
and this is sub-optimal for KD training. To keep the full distribution while being efficient, \cite{yang2022knowledge} approximated the distribution lattice with its one-best alignment, which was obtained while generating the transducer lattice and saved prior to training. KD then uses only on the one-best alignment instead of the whole lattice, leading to another approximation of the KD loss:
\begin{align}
    \mathcal{L}_{\text{KD}}\approx -\sum_{(t,u)\in \textsc{1Best}}\sum_{k=1}^{V} \mathbf{Z}_{\mathcal{T}}(t,u,k) \log \mathbf{Z}_{\mathcal{S}}(t,u,k).
    \label{eqn:1best alignment}
\end{align}
As the length of an alignment is $T+U$, this approximation reduces the computation from $\mathcal{O}(TUV)$ to $\mathcal{O}((T+U)V)$. When $V$ is large, a potential drawback of this method is that it would still use a large amount of memory. 

As a workaround, \cite{guo2022predicting} used the encoder features from an intermediate layer of a pre-trained teacher model for KD. Since the encoder output dimension of a teacher model is fixed, the computational cost is no longer related to $V$. To reduce the amount of data received from the teacher in on-the-fly KD, a multi-codebook vector quantisation method was used to compress 32-bit ``float'' features to 8-bit integers. During KD, the student model tries to predict the codebook indexes of teacher embeddings. This comes at a cost of slight performance degradation when compared to $L_1$ and $L_2$ loss \cite{guo2022predicting}.

Natural support for streaming is one reason why neural transducers are gaining more popularity. Streaming models have less future context and tend to emit non-blank symbols later than their non-streaming counterparts. Therefore, performing knowledge distillation on streaming transducers is more challenging when its teacher model is non-streaming. A 3-step distillation process is proposed in \cite{kurata20_interspeech} to delay the emission of the non-streaming teacher model using a streaming model before performing KD. Yang et al~\cite{yang2022knowledge} modified the 1-best alignment KD loss to improve the performance of the streaming student model. Directly applying \eqndot{1best alignment} to a streaming student could be problematic if the pre-trained teacher model is non-streaming as this would force the student to ``guess'' the future. Therefore, a time-shift variable $\tau$ was introduced to allow the student streaming model to emit symbols later than the teacher model. This leads to a modified version of \eqndot{1best alignment}:
\begin{align}
     \mathcal{L}_{\text{KD}}\approx -\sum_{(t,u)\in \textsc{1Best}}\sum_{k=1}^{K} \mathbf{Z}_{\mathcal{T}}(t,u,k) \log \mathbf{Z}_{\mathcal{S}}(t+\tau,u,k).
    \label{eqn:1best alignment streaming}
\end{align}

\subsection{Self-supervised learning for speech foundation models}
A foundation model refers to a large model pre-trained on a vast quantity of unlabelled data at scale based on SSL \cite{bommasani2021foundation}.   
It has become a new paradigm that an increasing number of research works are conducted by fine-tuning pre-trained foundation models using labelled data. 
Recently, many speech foundation models were developed and achieved state-of-the-art performance for various speech processing tasks \cite{baevski2020wav2vec, hsu2021hubert, chen2022wavlm}. Wav2vec 2.0 \cite{baevski2020wav2vec} used product quantisation to quantise acoustic features encoded by a stack of convolutional layers. The Gumbel softmax \cite{jang2016gumbelsoftmax} is applied to make the codebook selection process differentiable. SSL pre-training is carried out using a contrastive loss to encourage the Transformer encoder to discriminate each ground truth quantised vector from a set of distractors. HuBERT \cite{hsu2021hubert} applies k-means clustering to the speech representations and generates pseudo-class labels using the clustering results. During pre-training, the model is trained to predict the ground truth class labels of both masked and unmasked timestamps. To refine the clustering results, the features from an intermediate Transformer block are extracted for the second round of k-means clustering. WavLM \cite{chen2022wavlm} adopted the same idea as HuBERT for generating pseudo labels during pre-training, but used a more diverse pre-training dataset to increase the model's generalisation capability. It also performs speech denoising modelling during pre-training by adding noises and overlapping speech to the input. This not only enhances the robustness of the learned features, but also makes it more suitable for non-ASR tasks.

\subsection{Multi-teacher knowledge distillation}

In teacher-student training, the quality of the teacher model is crucial to the performance of the student model. Ensemble \cite{dietterich2000ensemble, rokach2010ensemble, lakshminarayanan2017simple} is a common approach for constructing stronger models by combining the output of different models. Lindqvist et al~\cite{ensembledistribution} used an ensemble as the teacher model to perform KD training, where a student model learns from the ensemble distribution obtained by averaging the distribution. However, the individual teacher information cannot be recovered from the weighted mean distribution, limiting the knowledge to be transferred from each information source. To address this information loss, \cite{fukuda2017efficient} proposed to randomly sample a teacher from a teacher pool for each training batch to so that the student model is exposed to individual teachers. In \cite{liu2020adaptive}, embeddings extracted from multiple teacher models are used as teacher targets and $L_2$ loss between the teacher embeddings and student embeddings was added as an auxiliary loss for training. Regarding ASR, \cite{JeremyDecisionTree} proposed an ensemble of hybrid teacher acoustic models constructed based on different tied-state triphones or tri-characters derived using the decision tree clustering approach.  
\section{Multi-Teacher Knowledge Distillation for Neural Transducers}
\label{sec:multi-teacher KD}

\subsection{Motivation}

The quality of the teacher model is a critical factor in KD training, since the resulting student model tends to have better performance with labels generated by better teacher models. Using an ensemble of multiple models is a common strategy to improve the teacher quality \cite{dietterich2000ensemble, sahraeian2018cross} and various attempts have been made to distil the knowledge from an ensemble of teacher models \cite{sau2016deep, zhao2020highlight}. 

SSL pre-training utilises the richness of unlabelled data and helps foundation models learn better feature representations. Good performance can be achieved when fine-tuning with even a small amount of labelled data. The data representations learned through different pre-training tasks could be very different, and maybe even complementary to each other. Therefore, it is desirable to use multiple SSL pre-trained foundation models to leverage their complementarity. In practice, it is reasonable to have multiple foundation models with similar performance, which can have different model structures or be trained on different datasets.
In this work, we propose a two-stage multi-teacher KD framework, where a student transducer model learns from multiple teacher foundation models at the same time. In the first stage, the encoder of the student transducer model is learned to regress the teacher models' embeddings. Extra unlabelled speech can be incorporated at this stage as no ground truth transcriptions are required. In the second stage, a neural transducer is initialised from the encoder in stage one, which will be then fine-tuned with audio-text pairs using the RNNT loss. Existing KD methods can be combined during the fine-tuning stage for performance improvement. 

\begin{figure*}
    \centering
    \includegraphics[width=0.95\textwidth]{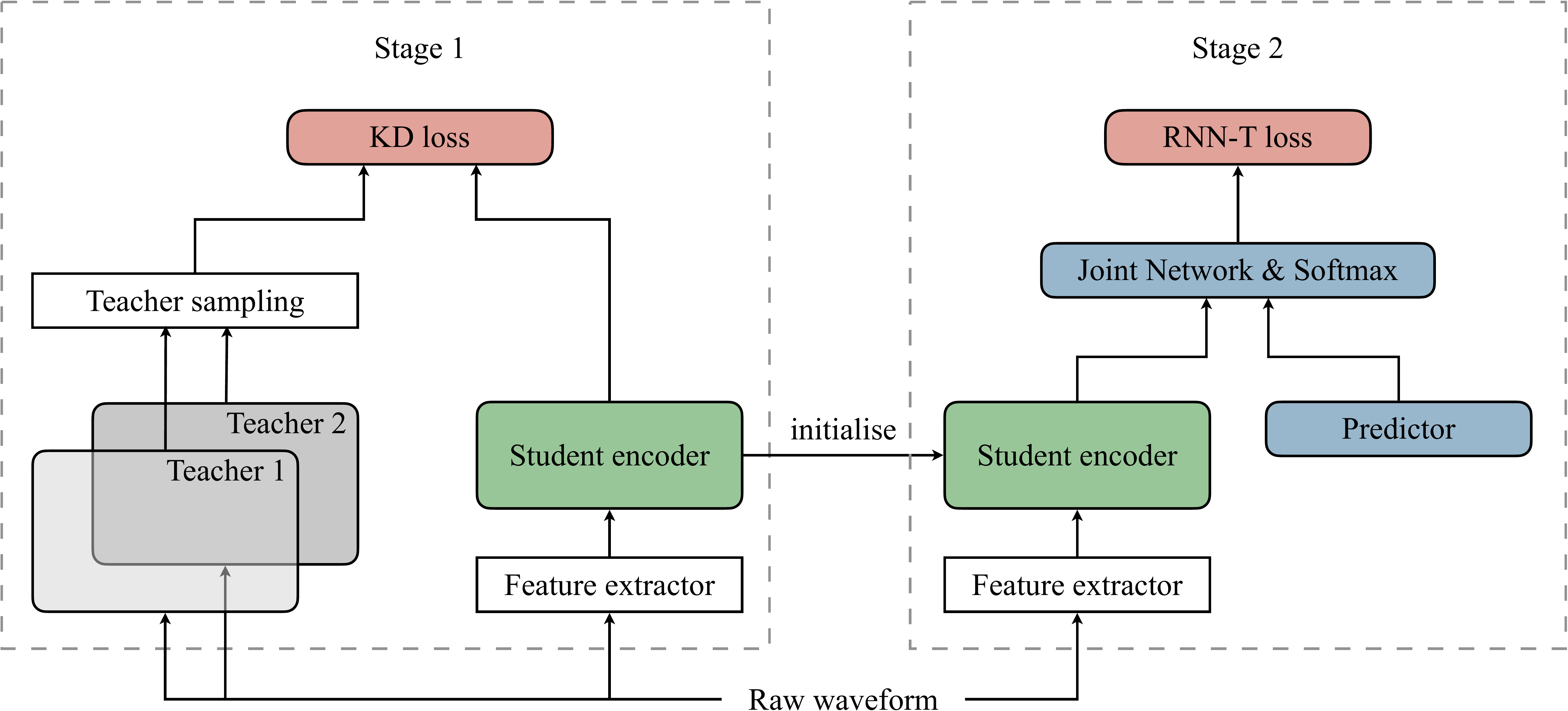}
    \caption{The proposed 2-stage KD framework. The left part describes the first stage: encoder pre-training. The teacher model receives raw waveform as input and generates embeddings. The right part is the second stage of training, where the student model is initialised from the well-trained encoder in the first stage and fine-tuned with audio-text pair.}
    \label{fig:framework}
\end{figure*}
\vspace{-0.2em}
\subsection{Encoder pre-training}
\label{sec:encoder pretraining}

The encoder of a neural transducer often has the majority of the total number of parameters. Therefore, we propose to carry out KD on encoder embeddings. The proposed KD training framework consists of two stages. In the first stage, only the encoder of the student model is considered while the decoder and joint network remain frozen. Assuming the number of teacher models to be $N$ and the encoder output dimension of each teacher to be $D^{\mathcal{T}_n}$. Given the same input acoustic features $\bm{X}$, the encoder of $n$-th teacher model generates $\mathcal{E}^{\mathcal{T}_n}=\bm{e}^{\mathcal{T}_n}_1,...,\bm{e}^{\mathcal{T}_n}_T$ as output. The encoder of the student model also generates embeddings $\mathcal{E}^{\mathcal{S}}=\bm{e}^{\mathcal{S}}_1,...,\bm{e}^{\mathcal{S}}_T$ of dimension $D^{\mathcal{S}}$. During training, the student attempts to regress the teacher embeddings $\mathcal{E}_n, n=1,...,N$ by sampling one teacher embedding for each training utterance in the mini-batch. Assuming the $n$-th teacher model is selected, the loss function for a training sample is as follows:
\begin{align}
    \mathcal{L}_{\text{KD}} = \frac{1}{T} \sum_{t=1}^T \textsc{Dist}(\textsc{LossNet}(\bm{e}^{\mathcal{S}_t}), \bm{e}^{\mathcal{T}_n}_t),
    \label{eqn:KD loss}
\end{align}
where $\textsc{LossNet}(\cdot)$ is a linear transformation that maps the student vector from $D^{\mathcal{S}}$ to $D^\mathcal{T}_n$ and $\textsc{Dist}(\cdot)$ is any function that measures the distance between two vectors of the same dimension, such as the commonly used $L_1$ and $L_2$ distances. Since KD is carried out at the encoder level, the computational and storage costs for encoder embedding level pre-training are independent of the output vocabulary size $V$, which alleviates the computation issue in distribution-based KD approaches for neural transducers\cite{panchapagesan2021efficient, yang2022knowledge}.

Streaming models tend to emit symbols later due to the lack of future context. Therefore, directly applying \eqndot{KD loss} to streaming student models can be problematic. Inspired by \cite{yang2022knowledge}, a time-delay factor $\tau$ is adopted during the first stage of training for streaming student models. The $t$-th frame of $n$-th teacher model's embedding will be aligned to the $\left(t+\tau \right)$-th frame of the student model. As a consequence, the last $\tau$ frames in the teacher model are discarded as no corresponding frames exist in the student model. 
This leads to a modified version of the distillation loss:
\begin{align}
    \mathcal{L}_{\text{KD}} = \frac{1}{T} \sum_{t=1}^{T-\tau} \textsc{Dist}(\textsc{LossNet}\left(\bm{e}^{\mathcal{S}}_t\right), \bm{e}^{\mathcal{T}_n}_{t+\tau}).
    \label{eqn:KD loss streaming}
\end{align}
Note that due to the introduction of the time-delay factor, the last $\tau$ frames of the student embedding have to be discarded during KD loss computation as no teacher embeddings exist for them. Intuitively, $\tau$ controls the trade-off between emission latency and model performance: a larger $\tau$ allows the student model to have more future contexts but increases emission delay and vice versa. However, a recent method applying this idea claims that the latency of the student model also decreases while the recognition accuracy also improves \cite{guo2022predicting}. With a carefully tuned $\tau$, this could be because the fixed time delay not only provides good supervision but also enforces the student model to emit earlier.

\subsection{Supervised fine-tuning}

In the second stage, the encoder pre-trained with embeddings from different teacher models in stage 1 is adopted as the initialisation of a student transducer model's encoder. The decoder and the joint network are randomly initialised. If the encoder is sufficiently trained, it should be able to accurately reproduce the output of the teacher model given the same input features. Therefore, this encoder can be treated as a weaker approximation to the self-supervised pre-trained model. The encoder-initialised student model is then fine-tuned with labelled speech to perform ASR tasks. To avoid catastrophic forgetting, a tri-stage learning rate schedule is used so that the model does not quickly depart from the initialisation. During the first warm-up stage, the learning rate increases linearly from a very small value to a warm-up learning rate. The learning rate keeps the same fixed value during the second stage and then decreases linearly to the final learning rate. The RNNT loss \cite{graves2012RNNT} is used to train the whole student model. Note that there is a fundamental difference between the proposed KD framework with existing embedding-level KD methods. Our framework splits the embedding learning and ASR training into two distinct processes whilst other methods perform KD in a multi-task fashion by treating KD as an auxiliary task during ASR training. An illustration of the proposed 2-stage KD framework is shown in \figdot{framework}

\section{Combination with other KD methods}

The proposed KD method performs distillation only using the output of the encoder, which does not involve the predictor and the joint network in the student neural transducer. Therefore, it is likely to be compatible with distribution-based KD methods as they focus on a different part of a transducer model. Here, we use the 1-best alignment KD \cite{yang2022knowledge} in addition to the proposed encoder embedding level KD. Under the single-teacher scenario, the 1-best loss can be added during supervised fine-tuning as an auxiliary loss to the normal RNNT loss. However, modifications need to be made when integrating the 1-best alignment-based KD under a multi-teacher KD setup. In this paper, we extend the 1-best alignment KD to n-best alignment KD by using the 1-best alignment from each individual teacher as the teacher label. By doing so, the loss function of n-best alignment KD becomes a weighted sum of $N$ 1-best KD loss defined in \eqndot{1best alignment}:
\begin{align}
    \mathcal{L}_{\text{nBest}} =  \sum_{n=1}^{N} \omega_n \mathcal{L}^{n}_{\text{1Best}}
\end{align}
where $\omega_n$ is the weight of the alignment decoded from the $n$-th teacher model and it holds $\sum_{n=1}^{N} \omega_n = 1$.

This auxiliary loss is interpolated with the original RNNT loss during the second stage of training, resulting in the final loss function:
\begin{align}
    \mathcal{L}_{\text{final}} = \mathcal{L}_{\text{RNNT}} + \lambda \mathcal{L}_{\text{nBest}},
\end{align}
where $\lambda$ is a tunable coefficient controlling the contribution of the auxiliary loss.
\section{Experimental Setup}
\label{sec:exp_setup}

\subsection{Model}

The Base version of Wav2vec 2.0 \cite{baevski2020wav2vec}, HuBERT \cite{hsu2021hubert} and WavLM \cite{chen2022wavlm} were selected as teacher models. For convenience, they will be referred to as ``W2v2'', ``HU'' and ``WL'' in the following experiments. Note that there are two Base versions of WavLM, with different amounts of unlabelled data used during pre-training. Here, the one pre-trained on MIX-94k hour is chosen as it is expected to have better performance. All three pre-trained models take the raw waveform as input and generate 50 encoder frames per second, whereas the most commonly used filter bank features for E2E trainable ASR usually operates at 25Hz. An extra subsampling module is appended to the pre-trained model, which concatenates two subsequent frames and applies a linear transform followed by a non-linear activation which eliminates the frame mismatch problem and the memory and computation cost is also reduced. The student model architecture is the small version of the Conformer transducer model (Cfm-S) \cite{gulati2020conformer} with 16 encoder layers of a hidden dimension of 144. To ensure the teacher model and student model have the same encoder feature dimension, an extra linear projection layer is added to the end of the encoder stack of the student model. All teacher models have a single-layer 640-d long short-term memory (LSTM) predictor network, and all student models have a single-layer 320-d LSTM predictor network. The streaming student model has zero future context. A triangular attention mask and causal convolution~\cite{oord2016wavenet} are used to ensure the model only attends to previous frames. The details of the teacher and student model and WERs of three teacher models are shown in \twotbl{model details}{teacher WER}. WavLM transducer achieves the lowest WERs among the three teachers because of the larger pretraining set. Note that the teacher models are about 10 times the size of the student model. All models are trained using the ESPnet~\cite{watanabe2018espnet} ASR toolkit.

\begin{table}[ht]
    \pretbl
    \begin{adjustbox}{width=\linewidth}
    \centering
    \begin{tabular}{c c c c c}
        \toprule
         & \multicolumn{3}{c}{Teacher Transducer} & \multicolumn{1}{c}{Student} \\
        \midrule
        Encoder & Wav2vec 2.0 \footnote{https://dl.fbaipublicfiles.com/fairseq/wav2vec/wav2vec\_small.pt} & HuBERT\footnote{https://dl.fbaipublicfiles.com/hubert/hubert\_base\_ls960.pt} & WavLM\footnote{https://github.com/microsoft/unilm/tree/master/wavlm} & Cfm-S \\
        Encoder size & 768-d & 768-d & 768-d & 144-d \\
        Pretraining data & LS-960 & LS-960 & MIX-94k & None \\
        Num. params & 99.2M & 99.5M & 99.2M & 10.3M \\
        \bottomrule
    \end{tabular}
    \end{adjustbox}
    \caption{Details of the teacher models. ``LS-960'' and ``MIX-94k'' denote the full LibriSpeech dataset and 94k mix dataset.}
    \label{tab:model details}
    \posttbl
    \posttbl
\end{table}

\begin{table}[!h]
    \centering
    \begin{tabular}{lcccc}
        \toprule
        \multirow{2}{*}{Teacher Transducer}    & \multicolumn{2}{c}{dev}   & \multicolumn{2}{c}{test}  \\
        \cmidrule(lr){2-3}  \cmidrule(lr){4-5}
        & \multicolumn{1}{c}{clean} & \multicolumn{1}{c}{other} & \multicolumn{1}{c}{clean} & \multicolumn{1}{c}{other} \\ 
        \midrule
        Wav2vec 2.0 (W2v2) & 5.1 & 12.2 & 5.2 & 11.8\\
        Hubert (Hu)  & 5.2 & 11.0 & 5.3 & 11.2 \\
        WavLM  (WL) & 3.9 & 8.4 & 3.9 & 8.3 \\
        \bottomrule
    \end{tabular}
    \caption{WERs of teacher models fine-tuned on LibriSpeech ``train-clean-100''. }
    \label{tab:teacher WER}
\end{table}

\subsection{Dataset}
The two popular datasets LibriSpeech~\cite{panayotov2015librispeech} and Libri-Light~\cite{kahn2020libri} were used for the experiments. The full LibriSpeech training set contains 960 hours of audiobook recordings with corresponding transcriptions and the LibriLight dataset contains 60+k hours of unlabelled speech. To verify the effectiveness of the proposed multi-teacher KD framework, experiments are first carried out on the ``train-clean-100" subset. All teacher models are also fine-tuned on the ``train-clean-100" subset using the raw waveform. During KD training, extra unlabelled speech is used to augment the training set for potential performance improvement. To investigate the effectiveness of this, the teacher model's embeddings of the remaining 860h of LibriSpeech and 1000 hours of speech randomly sampled from Libri-Light were extracted as teacher labels, resulting in a total pre-training dataset of around 2k hours. Pseudo transcriptions were also used to improve student models' performance. The pseudo transcriptions of 860 hours of remaining speech from LibriSpeech were obtained by performing beam search decoding of size 8 on individual teacher model. The transcriptions were not improved by language model fusion. 

During training, SpecAugment~\cite{park2019specaugment} was applied for data augmentation. Speed perturbation was not used in order to reduce the number of teacher labels that need to be stored. The output vocabulary has 256 sub-word units generated by SentencePiece~\cite{kudo2018sentencepiece}. During inference, the model with the lowest WER on the dev-other development set was chosen and the WERs on two test sets (test-clean and test-other) are reported.

\subsection{KD Training}
In the encoder pre-training stage, the raw waveform is fed to the fine-tuned teacher models to generate encoder embeddings. The masking in the teacher models was disabled and the embeddings were stored on disk prior to training. Various teacher model selection strategies were explored including uniform sampling and WER-related or similarity-related sampling. It is found that uniform sampling yields the best performance after fine-tuning and this sampling strategy is adopted in the rest of this paper. During training, the student model was given the 82-d filter bank features concatenated with 1-d pitch features corresponding to the waveform that is used to generate embeddings. SpecAugment is applied during student encoder pre-training to improve the robustness of features learned by the student encoder and Noam \cite{vaswani2017transformer} optimiser with warmup was used to update student encoder parameters. To evaluate the student model's encoder, the teacher embeddings of the ``dev-other" development set were also collected. The model with the lowest $L_1$ error on the ``dev-other" was selected as the initialisation for the second stage of training. When multiple teachers are used during KD, the averaged $L_1$ error over all teachers embeddings on ``dev-other'' was used as the model selection criterion.

During the second stage of KD training, the model was fine-tuned with audio-text pairs. The texts are either ground truth transcriptions (for ``train-clean-100'') or pseudo transcriptions decoded on unlabelled speech (the remaining 860-hour LibriSpeech). A tri-stage fine-tuning learning rate schedule was adopted, where the warm-up stage, constant stage and decay stage takes 10\%, 40\% and 50\% of the total training steps. The learning rate schedule for 100-hour and 960-hour experiments are shown in \tbl{lr schedule}.

\begin{table}[!h]
    \centering
    \begin{tabular}{c c c c c}
    \toprule
        Fine-tuning data    & Initial & Warmup & Decay \\
    \midrule
       100 hour  & 1e-6 & 1e-4 & 5e-6 \\
       960 hour  & 1e-6 & 5e-4 & 1e-5 \\ 
    \bottomrule
    \end{tabular}
    \caption{The learning rate schedule of fine-tuning in 100-hour and 960-hour experiments.}
    \label{tab:lr schedule}
\end{table}
\section{Experimental Results}
\label{sec:exp_results}
\subsection{Single Teacher vs Multi teacher}

\begin{table}[!ht]
    \centering
    \begin{adjustbox}{width=\linewidth}
    \begin{tabular}{lcccccc}
        \toprule
        \multirow{2}{*}{\shortstack[c]{Teacher\\model}} & \multicolumn{2}{c}{KD}  & \multicolumn{2}{c}{dev}   & \multicolumn{2}{c}{test}  \\
        \cmidrule(lr){2-3} \cmidrule(lr){4-5} \cmidrule(lr){6-7} 
         &\multicolumn{1}{c}{Embedding} & \multicolumn{1}{c}{nBest} & \multicolumn{1}{c}{clean} & \multicolumn{1}{c}{other} & \multicolumn{1}{c}{clean} & \multicolumn{1}{c}{other} \\ 
        \midrule
        \multicolumn{5}{l}{\textbf{Non-streaming baseline, no KD}}\\
        -  & - & - & 7.5 & 20.9 & 8.0 & 21.8 \\
        \midrule
        \multicolumn{5}{l}{\textbf{Single teacher KD}} \\
        W2v2 & \checkmark &  & 6.8 & 20.9 & 7.2 & 21.7 \\
        W2v2 & \checkmark & \checkmark & 6.6 & 20.7 & 6.8 & 20.9\\
        
        Hu& \checkmark & & 6.7 & 20.5 & 7.0 & 21.2 \\
        Hu& \checkmark & \checkmark & 6.6 & 20.3 & 6.8 & 20.7 \\
        
        WL & \checkmark &  & 6.5 & 20.2 & 6.7 & 20.7\\
        WL & \checkmark & \checkmark & \textbf{6.5} & \textbf{20.1} & \textbf{6.6} & \textbf{20.6} \\
        
        \midrule
        \multicolumn{5}{l}{\textbf{Multi teacher KD}}\\
        W2v2+Hu & & \checkmark & 6.6 & 19.0 & 6.9 & 19.4 \\
        W2v2+Hu & \checkmark &  & 6.4 & 19.6 & 6.6 & 19.8 \\
        W2v2+Hu & \checkmark & \checkmark  & 6.2 & 19.2 & 6.5 & 19.4\\
        
        WL+Hu & & \checkmark & 6.6 & 19.0 & 6.8 & 19.3 \\
        WL+Hu  & \checkmark & &6.2 & 19.2 & 6.6 & 19.5 \\
        WL+Hu & \checkmark & \checkmark & \textbf{6.1} & \textbf{18.9} & \textbf{6.5} & \textbf{19.2} \\
        
        \bottomrule
    \end{tabular}
    \end{adjustbox}
    \caption{\%WERs for non-streaming student transducer models. A tick under embedding KD or n-best KD means that embedding-level encoder pretraininng or 1-best (or n-best if multiple teachers are used) is adopted. ``W2v2'', ``Hu'' and ``WL'' stand for Wav2vec 2.0, HuBERT and WavLM respectively.}
    \label{tab:feature kd 100h}
\end{table}

Experiments were first carried out on the "train-clean-100" subset. Results on non-streaming transducers are reported in \tbl{feature kd 100h}. The following observations can be made. First, using encoder embeddings from self-supervised pre-trained models for KD successfully improves the performance of the student model. Second, the performance of the student model after embedding-level KD training is better when the teacher model has lower WERs. The best performing student model under single teacher setup is pre-trained using embeddings extracted from WavLM, which is also the best among the three teacher models, achieving 16.2\% and 5.0\% relative WERR on the test sets compared to the baseline model. Third, increasing the number of teachers leads to a further WER reduction. By using a teacher combination of HuBERT and WavLM during encoder pre-training, further relative WERRs of 1.5\% and 4.6\% is achieved compared to the best single-teacher setup. Fourth, embedding level KD is compatible with n-best alignment-based KD for both the single-teacher and multi-teacher setups. Larger WER reductions are observed after adding n-best alignment KD loss to the HuBERT+WavLM setup, leading to final relative WERRs of 18.8\% and 11.9\% compared to the baseline. Note that the performance improvement on test-clean is bigger than on the more challenging test-other set. This could be the reason that the student model is only pre-trained on the train-clean-100 subset, which is more acoustically similar to test-clean.

Experiments are then carried out for streaming student transducer models. Results in the 100-hour setup are shown in \tbl{feature kd streaming 100h}. To verify the effectiveness of $\tau$, student models trained with different $\tau$ values were compared under the single-teacher setup. It can be seen that $\tau=7$ consistently outperforms $\tau=0$ (which is equivalent to allowing no delay in the streaming student model), suggesting the necessity of choosing a sensible $\tau$ value during embedding level KD for streaming student model. The model trained with $\tau=0$ even performs worse than the baseline model trained without distillation. Therefore, $\tau$ is set to 7 in the following experiments. When combined with 1-best KD in the second stage, the same $\tau=7$ used during encoder pre-training was adopted when using \eqndot{1best alignment streaming} for one-best loss computation. Similar trends as in \tbl{feature kd 100h} can be observed for streaming student models. Using embedding-level KD consistently improves the student model under the single-teacher setup and a further gain is observed when two teachers are used for encoder pre-training or adding n-best alignment KD during fine-tuning. The best streaming student model is obtained under the two-teacher embedding KD setup with HuBERT and WavLM combined with 2-best KD during self-supervised fine-tuning, resulting in relative WERRs of 19.7\% and 7.8\% compared to the baseline model trained without KD. 


\begin{table}[!ht]
    \centering
    \begin{adjustbox}{width=\linewidth}
    \begin{tabular}{lcccccc}
        \toprule
        \multirow{2}{*}{\shortstack[c]{Teacher\\model}} & \multicolumn{2}{c}{KD}  & \multicolumn{2}{c}{dev}   & \multicolumn{2}{c}{test}  \\
        \cmidrule(lr){2-3} \cmidrule(lr){4-5} \cmidrule(lr){6-7} 
         &\multicolumn{1}{c}{Embedding} & \multicolumn{1}{c}{nBest} & \multicolumn{1}{c}{clean} & \multicolumn{1}{c}{other} & \multicolumn{1}{c}{clean} & \multicolumn{1}{c}{other} \\ 
        \midrule
        \multicolumn{5}{l}{\textbf{Streaming baseline, no KD}}\\
        - & - & - & 11.2 & 28.4 & 12.2 & 29.6 \\
        \midrule
        \multicolumn{5}{l}{\textbf{Single teacher KD}} \\
        W2v2, $\tau=0$ & \checkmark &  & 10.4 & 28.8 & 11.1 & 30.0 \\
        W2v2 & \checkmark &  & 9.9 & 27.4 & 10.2 & 28.5\\
        W2v2 & \checkmark & \checkmark & 9.9 & 27.4 & 10.1 & 28.4 \\
        
        Hu, $\tau=0$& \checkmark &  & 10.4 & 28.6 & 11.0 & 29.9 \\
        Hu &\checkmark & & 9.9 & 27.4 & 10.2 & 28.5 \\
        Hu &\checkmark & \checkmark & 9.9 & 27.1 & 10.1 & 28.3 \\
        
        WL, $\tau=0$ & \checkmark &  & 10.3 & 28.3 & 10.8 & 29.4\\
        WL & \checkmark &  & 9.8 & 27.2 & 10.1 & 28.2\\
        WL & \checkmark & \checkmark &  \textbf{9.7} & \textbf{26.9} & \textbf{10.1} & \textbf{28.0}  \\
        
        \midrule
        \multicolumn{5}{l}{\textbf{Multi teacher KD}}\\
        
        W2v2+Hu & \checkmark &  & 9.7 & 26.9 & 10.0 & 27.9 \\
        W2v2+Hu & \checkmark & \checkmark  & 9.6 & 26.8 & 9.9 & 27.7\\
        
        WL+Hu  & \checkmark & & 9.6 & 26.6 & 9.9 & 27.6 \\
        WL+Hu & \checkmark & \checkmark & \textbf{9.5} & \textbf{26.4} & \textbf{9.8} & \textbf{27.3} \\
        
        \bottomrule
    \end{tabular}
    \end{adjustbox}
    \caption{\%WERs for streaming student transducer models. $\tau$ is the time-shift factor used during embedding KD and is set to 7 unless explicitly specified.}
    \label{tab:feature kd streaming 100h}
\end{table}
\posttbl

\subsection{Effect of Unlabelled Data}

The effect of incorporating extra unlabelled data during encoder pretraining was investigated here. Experiments were first scaled up to include the remaining 860h audio from LibriSpeech as unlabelled data. The 960h audio features are fed to the teacher models for teacher embedding generation. The extracted features are utilised during encoder pre-training. Erroneous pseudo transcriptions obtained by decoding the teacher models on the 860 hour audio were used alongside the 100 hour ground truth transcriptions. This is expected to further improve the performance of the student model as reported in \cite{xu2020iterative, higuchi2021momentum}. The WERs of different teacher models decoded on the 860h data are shown in \tbl{WER unlabelled data}. 

\begin{table}[!ht]
    \centering
    \begin{tabular}{lcc}
    \toprule
    Teacher model  & clean-360 & other-500\\
    \midrule
    W2v2 & 4.9  & 6.4\\
    Hu     & 5.0 & 6.2 \\
    WL     & 4.0 & 5.2\\
    \bottomrule
    \end{tabular}
    \caption{\%WERs of teacher models on train-clean-360 and train-other-500 subset. }
    \label{tab:WER unlabelled data}
\end{table}


The WERs of non-streaming student models trained on 960h data are shown in \tbl{feature kd 960h}. The WERs of models trained using only pseudo transcriptions (alongside with 100h ground truth) without KD are also listed for comparison (see section ``No KD'' in \tbl{feature kd 960h}). The following points could be drawn. First, the student model's performance is significantly improved by using only the pseudo transcriptions. The better the pseudo transcriptions are, the lower WERs the student model achieves. Second, the student model consistently benefits from single teacher embedding-level KD, resulting in an averaged relative WERR of 9.6\% and 10.9\% on three teacher models. Third, the performance of the student model is further improved under the multi-teacher encoder pre-training setup, achieving further relative WERR of 4.7\% and 2.2\% with WavLM and HuBERT compared to the best single-teacher setup. Finally, slight performance improvement is achieved when n-best alignment KD is also used under a multi-teacher setup. 

        
        
        
        
        
        

\begin{table}[!ht]
    \centering
    \begin{adjustbox}{width=\linewidth}
    \begin{tabular}{lcccccc}
        \toprule
        \multirow{2}{*}{\shortstack[c]{Teacher\\model}} & \multicolumn{2}{c}{KD}  & \multicolumn{2}{c}{dev}   & \multicolumn{2}{c}{test}  \\
        \cmidrule(lr){2-3} \cmidrule(lr){4-5} \cmidrule(lr){6-7} 
         &\multicolumn{1}{c}{Embedding} & \multicolumn{1}{c}{nBest} & \multicolumn{1}{c}{clean} & \multicolumn{1}{c}{other} & \multicolumn{1}{c}{clean} & \multicolumn{1}{c}{other} \\ 
        \midrule
        \multicolumn{5}{l}{\textbf{No KD}}\\
        Baseline & - & - & 7.5 & 20.9 & 8.0 & 21.8\\
        W2v2 trans & - & - & 5.3 & 11.3 & 5.4 & 11.6 \\
        Hu trans & - & - & 5.1 & 10.6 & 5.3 & 10.8 \\
        WL trans & - & - & 4.4 & 10.2 & 4.6 & 10.2 \\
        \midrule
        \multicolumn{5}{l}{\textbf{Single teacher KD}} \\
        W2v2 & \checkmark &  & 4.7 & 9.8 & 4.7 & 9.9\\
        
        Hu& \checkmark &  & 4.7 & 9.6 & 4.8 & 9.8  \\
        
        WL & \checkmark &  & \textbf{4.2} & \textbf{9.2} & \textbf{4.3} & \textbf{9.3} \\
        
        \midrule
        \multicolumn{5}{l}{\textbf{Multi teacher KD}}\\
        W2v2+Hu & \checkmark &  & 4.5 & 9.4 & 4.5 & 9.6 \\
        W2v2+Hu & \checkmark & \checkmark  & 4.4 & 9.2 & 4.4 & 9.5\\

        WL+Hu  & \checkmark & & 3.9 & 9.0 & 4.1 & 9.1 \\
        WL+Hu & \checkmark & \checkmark & \textbf{3.9} & \textbf{8.9} & \textbf{4.0} & \textbf{9.0}  \\
        
        \bottomrule
    \end{tabular}
    \end{adjustbox}
    \caption{\%WERs for non-streaming student transducer models trained on 960h data. ``trans'' denotes the 860 hour pseudo transcription obtained from teacher model using beam search decoding. }
    \label{tab:feature kd 960h}
\end{table}

\begin{table}[!ht]
    \centering
    \begin{adjustbox}{width=\linewidth}
    \begin{tabular}{lcccccc}
        \toprule
        \multirow{2}{*}{\shortstack[c]{Teacher\\model}} & \multicolumn{2}{c}{KD}  & \multicolumn{2}{c}{dev}   & \multicolumn{2}{c}{test}  \\
        \cmidrule(lr){2-3} \cmidrule(lr){4-5} \cmidrule(lr){6-7} 
         &\multicolumn{1}{c}{Embedding} & \multicolumn{1}{c}{nBest} & \multicolumn{1}{c}{clean} & \multicolumn{1}{c}{other} & \multicolumn{1}{c}{clean} & \multicolumn{1}{c}{other} \\ 
        \midrule
        \multicolumn{5}{l}{\textbf{No KD}}\\
        Baseline  & - & -  & 11.2 & 28.4 & 12.2 & 29.6 \\
        W2v2 trans & - & - & 9.5 & 18.9 & 10.6 & 19.4 \\
        Hu trans & - & - & 9.3 & 17.5 & 10.1 & 17.8 \\
        WL trans & - & - & 8.4 & 15.9 & 9.5 & 16.2\\
        \midrule
        \multicolumn{5}{l}{\textbf{Single teacher KD}} \\
        W2v2 & \checkmark &  & 7.6 & 16.1 & 8.1 & 17.0\\
        
        Hu& \checkmark &  & 7.3 & 15.5 & 7.3 & 15.7   \\
        
        WL & \checkmark &  & \textbf{6.8} & \textbf{15.3} & \textbf{7.2} & \textbf{15.3} \\
        
        \midrule
        \multicolumn{5}{l}{\textbf{Multi teacher KD}}\\
        W2v2+Hu & \checkmark &  & 7.1 & 15.3 & 7.1 & 15.6 \\
        W2v2+Hu & \checkmark & \checkmark  & 7.0 & 15.1 & 7.1 & 15.4 \\

        WL+Hu  & \checkmark & & 6.7 & 15.0 & 7.0 & 15.2 \\
        WL+Hu & \checkmark & \checkmark & \textbf{6.6} & \textbf{14.9} & \textbf{6.8} & \textbf{14.9}  \\
        
        \bottomrule
    \end{tabular}
    \end{adjustbox}
    \caption{\%WERs for streaming student transducer models trained on 960h data. ``trans'' denotes the pseudo transcription obtained from teacher model using beam search decoding. $\tau$ is set to 7 in all KD experiments.}
    \label{tab:streaming feature kd 960h}
\end{table}

The WERs of streaming models are shown in \tbl{streaming feature kd 960h}. Here, $\tau$ is fixed to 7 as it produces the best results in \cite{yang2022knowledge}. Again, the models trained using only pseudo transcriptions from the teacher models are shown in the first section in \tbl{streaming feature kd 960h} for comparison. The same trend of performance improvement can be observed for streaming models when multi-teacher KD was applied. When using HuBERT and WavLM jointly during encoder pre-training, a WERR of 40.1\% and 48.6\% was achieved compared to the baseline student model trained without KD.

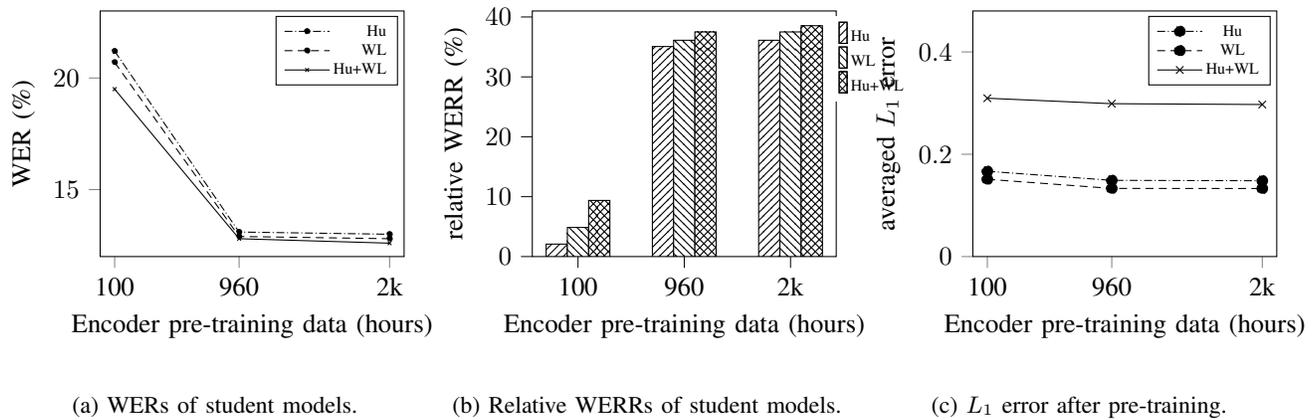
\begin{figure*}[!ht]
    \centering
    \begin{subfigure}[b]{.31\textwidth}
        \centering
        \begin{tikzpicture}
\begin{axis}[
    legend cell align={right},
    legend style={nodes={scale=0.6, transform shape}},
    width=\textwidth,
    tick align=outside,
    tick pos=left,
    xlabel=Encoder pre-training data (hours),
    ylabel= WER (\%),
    xmin=0, xmax=2100,
    ymin=12, ymax=23,
    xtick={100,960, 2000},
    xticklabels={100,960, 2k},   
    ytick={10,4,...,23}
            ]
\addplot[mark=*,black, mark size=1, densely dashdotted] 
plot coordinates {
    (100,21.2)
    (960,13.1)
    (2000,13.0)
};
\addlegendentry{Hu}

\addplot[color=black,mark=*, mark size=1, densely dashed]
    plot coordinates {
        (100,20.7)
        (960,12.9)
        (2000,12.8)
    };
\addlegendentry{WL}

\addplot[color=black,mark=x, mark size=1]
    plot coordinates {
        (100,19.5)
        (960,12.8)
        (2000,12.6)
    };
\addlegendentry{Hu+WL}
\end{axis}
\end{tikzpicture}
        \caption{WERs of student models.}
        \label{fig:WER pretraining 100h}
    \end{subfigure}
    \begin{subfigure}[b]{.31\textwidth}
    \centering

\pgfplotsset{compat=1.11,
    /pgfplots/ybar legend/.style={
    /pgfplots/legend image code/.code={%
       \draw[##1,/tikz/.cd,yshift=-0.25em]
        (0cm,0cm) rectangle (3pt,0.8em);},
   },
}
\begin{tikzpicture}
\definecolor{darkgray176}{RGB}{176,176,176}
\begin{axis}[
legend cell align={left},
legend style={
    nodes={scale=0.6, transform shape},
   ,
  anchor=north west,
  draw=none
},
width=\textwidth,
tick align=outside,
tick pos=left,
x grid style={darkgray176},
xlabel={Encoder pre-training data (hours)},
xmin=-0.23, xmax=2.63,
xtick style={color=black},
xtick={0.2,1.2,2.2},
xticklabels={100,960,2k},
y grid style={darkgray176},
ylabel={relative WERR (\%)},
ymin=0, ymax=41,
ytick style={color=black},
]
\draw[draw=black,fill=white,postaction={pattern=north east lines}] (axis cs:-0.1,0) rectangle (axis cs:0.1,2.08333333333334);
\draw[draw=black,fill=white,postaction={pattern=north east lines}] (axis cs:0.9,0) rectangle (axis cs:1.1,35.0694444444444);
\draw[draw=black,fill=white,postaction={pattern=north east lines}] (axis cs:1.9,0) rectangle (axis cs:2.1,36.1111111111111);
\addlegendimage{ybar,ybar legend,draw=black,fill=white,postaction={pattern=north east lines}}
\addlegendentry{Hu}

\draw[draw=black,fill=white,postaction={pattern=north west lines}] (axis cs:1.1,0) rectangle (axis cs:1.3,36.1111111111111);
\draw[draw=black,fill=white,postaction={pattern=north west lines}] (axis cs:2.1,0) rectangle (axis cs:2.3,37.5);
\draw[draw=black,fill=white,postaction={pattern=north west lines}] (axis cs:0.1,0) rectangle (axis cs:0.3,4.86111111111112);
\addlegendimage{ybar,ybar legend,draw=black,fill=white,postaction={pattern=north west lines}}
\addlegendentry{WL}

\draw[draw=black,fill=white,postaction={pattern=crosshatch}] (axis cs:1.3,0) rectangle (axis cs:1.5,37.5);
\draw[draw=black,fill=white,postaction={pattern=crosshatch}] (axis cs:2.3,0) rectangle (axis cs:2.5,38.5416666666667);
\draw[draw=black,fill=white,postaction={pattern=crosshatch}] (axis cs:0.3,0) rectangle (axis cs:0.5,9.375);
\addlegendimage{ybar,ybar legend,draw=black,fill=white,postaction={pattern=crosshatch}}
\addlegendentry{Hu+WL}

\end{axis}
\end{tikzpicture}
    \caption{Relative WERRs of student models.}
    \label{fig:WERR pretraining 100h}
    \end{subfigure}
    \begin{subfigure}[b]{.31\textwidth}
    \centering

\begin{tikzpicture}
\begin{axis}[
    legend style={nodes={scale=0.6, transform shape}},
    width=\textwidth,
    tick align=outside,
    tick pos=left,
    xlabel=Encoder pre-training data (hours),
    ylabel=averaged $L_1$ error,
    xmin=0, xmax=2100,
    ymin=0, ymax=0.48,
    xtick={100,960, 2000},
    xticklabels={100,960, 2k},   
    ytick={0.1,0.05,...,0.4}
            ]
\addplot[mark=*,black, densely dashdotted] 
plot coordinates {
    (100,0.1667)
    (960,0.1490)
    (2000,0.1482)
};
\addlegendentry{Hu}

\addplot[color=black,mark=*, densely dashed]
    plot coordinates {
        (100,0.1511)
    (960,0.1332)
    (2000,0.13302)
    };
\addlegendentry{WL}

\addplot[color=black,mark=x]
    plot coordinates {
        (100,0.3095)
    (960,0.2988)
    (2000,0.2972)
    };
\addlegendentry{Hu+WL}
\end{axis}
\end{tikzpicture}
    \caption{$L_1$ error after pre-training.}
    \label{fig:l1 error}
    \end{subfigure}
    \caption{\%WERs and relative WERRs of student models pre-trained with different amount of acoustic data. All models were fine-tuned with the 100 hour ground truth data.}
    
\end{figure*}

\begin{table}[!ht]
    \centering
    \begin{tabular}{lcccc}
        \toprule
        \multirow{2}{*}{Teacher model}    & \multicolumn{2}{c}{dev}   & \multicolumn{2}{c}{test}  \\
        \cmidrule(lr){2-3}  \cmidrule(lr){4-5}
        & \multicolumn{1}{c}{clean} & \multicolumn{1}{c}{other} & \multicolumn{1}{c}{clean} & \multicolumn{1}{c}{other} \\ 
        \midrule
        \multicolumn{5}{l}{\textbf{100h fine-tune}}\\

        Hu  & 7.3 & 15.5 & 7.3 & 15.7 \\
        
        WL  & 6.8 & 15.3 & 7.2 & 15.3 \\
        
        WL+Hu  & 6.8 & 15.0 & 7.0 & 15.2 \\
        \midrule
        \multicolumn{5}{l}{\textbf{960h fine-tune}}\\
        
        Hu & 4.6 & 9.4 & 4.6 & 9.7 \\
        WL & 4.1 & 9.2 & 4.2 & 9.2 \\
        WL+Hu  & 3.8 & 8.8 & 3.9 & 9.0 \\
        
        \bottomrule
    \end{tabular}
    \caption{\%WERs for non-streaming student transducer models pre-trained on 2k hours. Only embedding KD is adopted.}
    \label{tab:extra unlabelled}
\end{table}

To investigate if extra unlabelled data from another domain helps embedding-level KD, the extra 1000 hours of unlabelled data from Libri-Light was incorporated during encoder KD training for non-streaming student models. Fine-tuning with train-clean-100 ground truth data and 960 hours mix data are carried out and the WERs are shown in \tbl{extra unlabelled}. 
Compared to the result in \tbl{feature kd 960h}, the WERs were further improved after increasing the pre-training dataset from 960 to 2k hours.

\subsection{Pretraining Error and Finetuning Accuracy}

The previous experiments demonstrated that the student model achieves better performance with the increase of pre-training data, indicating that a better encoder initialisation is learned in the first stage. Apart from the WER after fine-tuning, the KD training in the first stage can also be evaluated with the $L_1$ error on test sets. Here, the relationship between the $L_1$ error after pre-training and the final WER is investigated. The fine-tuning results of the student models pre-trained with 100 hours, 960 hours and 2k hours data are shown in \figdot{WER pretraining 100h} and \figdot{WERR pretraining 100h}. Note that only train-clean-100 is used during fine-tuning.

The normalised $L_1$ error on the development set of student models pre-trained with different amounts of unlabelled data is illustrated in \figdot{l1 error}. The $L_1$ error decreases with extra unlabelled data used during encoder pre-training for both the single-teacher and multi-teacher setups. This is in-line with the improved student model performance after fine-tuning. However, a lower $L_1$ error does not guarantee a better student model after fine-tuning when comparing single-teacher and multi-teacher setups (see \figdot{WERR pretraining 100h}). The student model trained with two teachers (HuBERT and WavLM) has a much higher $L_1$ error than the single-teacher setup, but yields lower WERs after fine-tuning. This could be the reason that the random teacher model selection strategy drives the encoder initialisation to a manifold that is in between the teacher models and can be quickly adapted to the fine-tuning data. This also effectively prevents the student model from overfitting to a single teacher, which is beneficial for multi-teacher training.



        
        

\section{Conclusions}
\label{sec:conclusions}
In this paper, a two-stage teacher-student framework has been proposed, where a student neural transducer ASR model distils the knowledge either from one or multiple complementary SSL pre-trained speech foundation models. In the first stage, the student ASR encoder is trained to approximate the embeddings generated by one or multiple teacher encoders without using the ground truth labels. In the second stage, the entire student model is fine-tuned with paired audio-text data, where the paired texts can be generated either by human annotation or by existing teacher ASR models. On LibriSpeech 100h, the averaged WER of a non-streaming student model trained with a single teacher is 11\% relative lower than that trained from scratch and using the multi-teacher setup further increases the relative WER reduction to 14\%. The proposed KD framework is also effective for streaming neural transducers using an additional time-delay factor, which is to resolve the emission mismatch between the non-streaming teacher and the streaming student. The proposed KD framework is also complementary to existing KD methods, leading to further performance improvement in combination. Further WER reductions can be achieved when scaling up the amount of unlabelled data used in the first stage. The best-performing student is obtained under a multi-teacher setup with extra unlabelled data, resulting relative WERR of 55\%.

\bibliographystyle{IEEEtran}
\bibliography{ref.bib}

\end{document}